\def\year{2021}\relax
\documentclass[letterpaper]{article} 
\usepackage{aaai21}  
\usepackage{times}  
\usepackage{helvet} 
\usepackage{courier}  
\usepackage[hyphens]{url}  
\usepackage{graphicx} 
\usepackage{natbib}  
\usepackage{caption} 
\usepackage{booktabs} 
\usepackage{physics}
\usepackage{subfig}

\title{Identifying Influential Brokers on Social Media from Social
Network Structure}

\author{
    Sho Tsugawa, \textsuperscript{\rm 1}
    Kohei Watabe\textsuperscript{\rm 2}
    \\
}
\affiliations{
    \textsuperscript{\rm 1}Faculty of Engineering, Information and Systems, University of Tsukuba\\
Tsukuba, Ibaraki 305--8573, Japan\\
    \textsuperscript{\rm 2}Graduate School of Engineering, Nagaoka University of Technology\\
Nagaoka, Niigata 940--2188, Japan\\

    s-tugawa@cs.tsukuba.ac.jp, k\_watabe@vos.nagaokaut.ac.jp

}

\begin{document}

\maketitle

\begin{abstract}
Identifying influencers in a given social network has become an important
 research problem for various applications, including accelerating
 the spread of information in viral marketing and preventing the spread of  fake news and rumors.  The literature contains a rich body of
 studies on identifying influential {\em source spreaders} who can
 spread their {\em own} messages to many other nodes.  In contrast,
 the identification of influential {\em brokers} who can spread {\em other
 nodes'} messages to many nodes has not been fully explored.
 Theoretical and empirical studies suggest that involvement of both
 influential source spreaders and brokers is a key to facilitating
 large-scale information diffusion cascades.  Therefore, this paper 
 explores ways to identify influential brokers from a given
 social network.  By using three social media datasets, we investigate
 the characteristics of influential brokers by comparing them with
 influential source spreaders and central nodes obtained from centrality
 measures.  Our results show that (i) most of the influential source
 spreaders are not influential brokers (and vice versa) and (ii) the
 overlap between central nodes and influential brokers is small (less
 than 15\%) in Twitter datasets.  We also tackle the problem of
 identifying influential brokers from centrality measures and node
 embeddings, and we examine the effectiveness of social network features in
 the broker identification task.  Our results show that (iii) although a
 single centrality measure cannot characterize influential
 brokers well, prediction models using node
 embedding features achieve F$_1$ scores of 0.35--0.68,
 suggesting the effectiveness of social network features for identifying
 influential brokers.

\end{abstract}

\section{Introduction}
\label{sec:intro}

Identifying influencers from a social network has been a fundamental
research task in the web and network science research
communities~\cite{lu2016vital,Li18:Influence,Morone15:CI,al2018analysis,banerjee2020survey}.
It has been shown that a few individuals called influencers play an
important role in triggering a large-scale cascade of information
diffusion~\cite{pei2014,Katz55:personal}.  Thus, identifying influencers
is considered to be crucial for conducting effective viral marketing
campaigns~\cite{Domingos02:Mining,Domingos01:Mining,Kempe03:Maximizing}
and preventing the spread of unwanted information (e.g., fake news and
rumors)~\cite{Budak11:Limiting}.

Several algorithms for identifying influencers have been
proposed~\cite{lu2016vital,Li18:Influence,al2018analysis,banerjee2020survey}.  A common
approach is to calculate centrality measures of nodes in a social
network and extract the nodes with high centrality as
influencers~\cite{chen2012,lu2016vital,Morone15:CI}.  Traditional centrality measures include 
degree~\cite{Freeman79:Centrality},
closeness~\cite{Freeman79:Centrality},
betweenness~\cite{Freeman79:Centrality},
PageRank~\cite{Brin98:PageRank}, and $k$-core
index~\cite{seidman1983,Dorogovtsev06:k}. New measures have been also used
for identifying influencers, including
VoteRank~\cite{zhang2016identifying} and Collective
Influence (CI)~\cite{Morone15:CI}.  More recently,
combining multiple centrality measures in a machine-learning framework
has been
proposed for identifying influencers~\cite{bucur2020top,zhao2020machine}.

\begin{figure}[t]
  \centering 
\includegraphics[clip,width=.74\columnwidth]{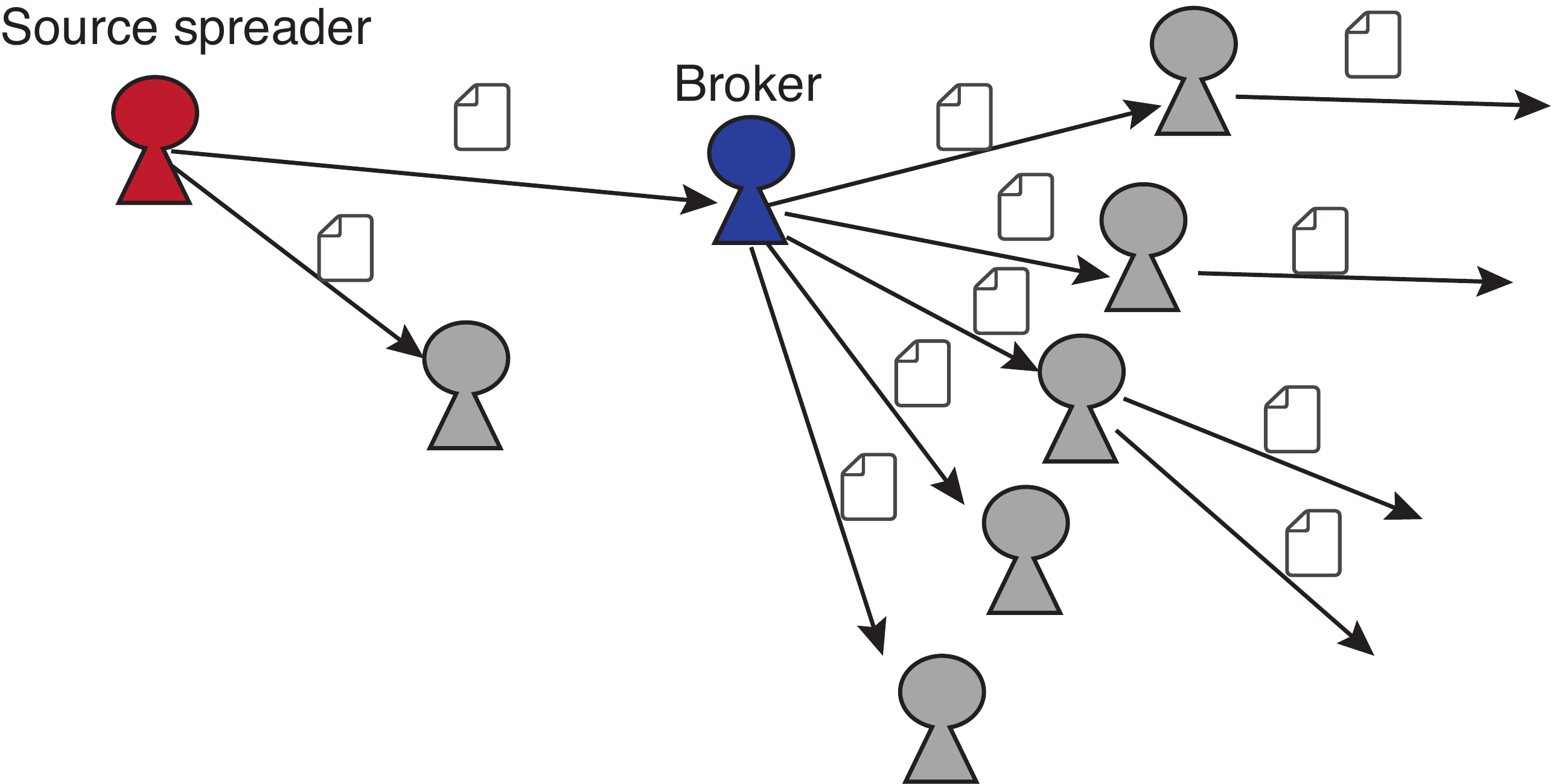}  
\caption{Source spreaders and brokers. Influential source spreaders are users who can spread their own messages to many other users, and influential brokers are users who can spread other users' messages to many users.}
  \label{fig:source-intermediate}
\end{figure}

While most previous studies assumed either explicitly or implicitly that
influencers are nodes who can spread their {\em own} messages to many
other nodes~\cite{Kempe03:Maximizing,Li18:Influence,zhang2016identifying,lu2016vital,chen2012,banerjee2020survey}, another type of influencer on social media who
can spread {\em other users'} messages to many users has also been shown to
play an important role in large-scale information
diffusion~\cite{Bakshy12:Role,Weng13:Virality,liu2012rising,araujo2017getting,meng2018diffusion,tsugawa2019empirical}.
We refer to the former as {\em source spreaders} and the latter as {\em
brokers}~\cite{burt2000network,araujo2017getting,meng2018diffusion} (Fig.~\ref{fig:source-intermediate}).  Bakshy et
al.~\shortcite{Bakshy12:Role}, and Weng et al.~\shortcite{Weng13:Virality} showed
that information diffusion by brokers who bridge different communities
affects the final sizes of information diffusion cascades.  Araujo et
al.~\shortcite{araujo2017getting} showed that information brokers facilitate
content diffusion for global brands. Meng et al.~\shortcite{meng2018diffusion}
showed that the diffusion size of health-related information posted by
the Centers for Disease Control and Prevention is significantly
affected by broker involvement.  The importance of brokers has also been
discussed in the context of social capital theory~\cite{burt2000network}.
Therefore, identifying influential brokers, as well as source spreaders,
is important for successful information dissemination.
For example, if a company wishes to disseminate the tweets of
its official account, then influential brokers are helpful for 
doing so widely. 
In another scenario, influential brokers may unintentionally disseminate 
unwanted information such as fake news posted by troll accounts.
In that case, it
would be effective to identify those influential brokers in order to limit the
spread of unwanted information.

In this paper, we aim to understand the characteristics of influential brokers and identify them from a given social network among social media users.
In particular, we address the following research questions.
\begin{itemize}
 \item {\bf (RQ1)} How different are influential source spreaders and influential brokers?
 \item {\bf (RQ2)} Are influential brokers located at central positions in a social network?
 \item {\bf (RQ3)}  How accurately can we predict the influential brokers from a social network
       by using node embeddings, which incorporate complex structural
       information about the social network?

\end{itemize}
Because the characteristics of influential brokers have been unclear to date, we
begin by examining their characteristics by comparison
with influential source spreaders ({\bf RQ1}) and central nodes
obtained from centrality measures ({\bf RQ2}). We then conduct
experiments to identify influential brokers.  We focus on node
embeddings~\cite{Rossi18:TKDE,Rossi18:WWW,goyal2018graph,grover2016node2vec,cui2018survey,qiu2018network,qiu2019netsmf},
which are low-dimensional vector representations of nodes, and
traditional centrality
measures~\cite{Freeman79:Centrality,Brin98:PageRank,seidman1983,Dorogovtsev06:k}
as features for identifying influencers, and we examine their effectiveness
({\bf RQ3}).

Our main contributions are summarized as follows.
\begin{itemize}

 \item We tackle the problem of identifying influential brokers from a
       social network. Although previous empirical results~\cite{Bakshy12:Role,Weng13:Virality,liu2012rising,araujo2017getting,meng2018diffusion,tsugawa2019empirical}
       suggested the importance of brokers for triggering a large-scale
       cascade, the problem of identifying influential brokers has rarely been
       explored.

 \item We empirically show the effectiveness of node embeddings as
       well as the limitations of using a single traditional centrality measure for
       identifying influential brokers.  Node embeddings have been shown
       to be effective for several tasks including link prediction~\cite{grover2016node2vec,Rossi18:TKDE} and
       node label classification~\cite{grover2016node2vec,qiu2018network,qiu2019netsmf}, and we show their
       effectiveness in influencer identification tasks.

 \item We examine the characteristics of influential source spreaders and
       brokers in different domains.  We use three different types of
       datasets that come from different domains with different
       languages, and we aim to understand whether there are universal
       characteristics of influencers among different domains.
\end{itemize}

\section{Related Work}
\label{sec:related} 

The literature contains multiple definitions of influential or
important social media users~\cite{Riquelme16:Measuring}.  For instance,
Cha et al.~\shortcite{Cha10:Measuring} examined the influence of Twitter users
using their numbers of followers, retweets, and 
mentions, and Bakshy et al.~\shortcite{Bakshy11:Everyone's} considered users
who initiate large-scale retweet cascades as being influencers.  Influencers
are often referred to by other terms such as opinion
leaders~\cite{hu2012breaking} and
authorities~\cite{Bouguessa15:Identifying}.  These studies regard
influencers as users who can spread their own information or posts to
many other users, and we refer to this type of influencer as an influential
source spreader.  Another line of research~\cite{araujo2017getting,liu2012rising,li2014social,burt2000network}
is focused on influential brokers who can spread other users' information or
posts to many users.  Information disseminated by influential brokers
who bridge different communities is suggested to spread widely in
several domains such as YouTube videos~\cite{liu2012rising},
brand content~\cite{araujo2017getting}, and health-related
information~\cite{meng2018diffusion}.  Araujo et
al.~\shortcite{araujo2017getting} and Li et al.~\shortcite{li2014social} showed that
both influential source spreaders and brokers play important
roles in facilitating large-scale information diffusion cascades.

Although there are multiple definitions of influencers, most algorithms
for identifying influencers aim either explicitly or implicitly to identify
influential source
spreaders~\cite{Kempe03:Maximizing,Li18:Influence,banerjee2020survey,zhang2016identifying,pei2014,chen2012}.
A common way to evaluate the effectiveness of influencer
identification algorithms is to use synthetic information diffusion
models such as the susceptible--infected--removed (SIR) model~\cite{zhang2016identifying,Li14:Identifying,chen2012}, the independent
cascade model~\cite{Kempe03:Maximizing}, and the linear threshold
model~\cite{Kempe03:Maximizing}.  As a metric for evaluating the
effectiveness of the algorithms, the number of users who receive
information when the identified influencers are selected as {\em seed}
nodes in the information diffusion models is
used~\cite{Kempe03:Maximizing,zhang2016identifying,Li14:Identifying}.  In other words, many
existing algorithms are shown to be effective for identifying users who
can spread information to many other users under synthetic
information diffusion models.  Some
studies~\cite{pei2014,panagopoulos2020influence,tsugawa2018identifying}
used real information diffusion trace data rather than synthetic models,
but those studies also evaluated the users' power as influential
source spreaders.

Algorithms for identifying influential source spreaders fall roughly
into three categories: (i) algorithms based on network
topology~\cite{Morone15:CI,zhang2016identifying,lu2011leaders,Li14:Identifying,chen2012},
(ii) algorithms based on information diffusion
models~\cite{Kempe03:Maximizing,Li18:Influence,banerjee2020survey,Tang15:Influence,Tang14:TIM}, and
(iii) algorithms based on the records of users'
activities~\cite{weng2010,yamaguchi2010turank,panagopoulos2020influence}.  Algorithms based on network topology
estimate the influence of each user from the topological structure of a
social network using network metrics of nodes. The metrics include
traditional centrality
measures~\cite{Freeman79:Centrality,Brin98:PageRank},
LeaderRank~\cite{lu2011leaders}, VoteRank~\cite{zhang2016identifying},
and CI~\cite{Morone15:CI}.  The second category of algorithms, namely, those that
use information diffusion models, are referred to as influence maximization
algorithms~\cite{Kempe03:Maximizing,Li18:Influence,banerjee2020survey};
these identify a set of influential seed
nodes that can spread information to many nodes under the given
information cascade
model~\cite{Kempe03:Maximizing,Li18:Influence,banerjee2020survey}.  In
the seminal work by Kempe et al.~\shortcite{Kempe03:Maximizing}, the
influence maximization problem was formulated as a combinatorial
optimization problem, and greedy algorithms were proposed.  Since then,
several efficient influence maximization
algorithms have been proposed, including TIM~\cite{Tang14:TIM} and
IMM~\cite{Tang15:Influence} that can find influencers in huge networks
based on influence cascade models.  While the first two categories of 
algorithms use only social network structure and synthetic diffusion
models, the third category of algorithms uses the records of users'
activities such as tweets and retweets in addition to the network
structure.  Such algorithms that are widely used include TwitterRank~\cite{weng2010},
TURank~\cite{yamaguchi2010turank}, and
CELFIE~\cite{panagopoulos2020influence}.  These algorithms have been shown to
be effective for identifying influential source spreaders, but their
effectiveness for identifying influential brokers remains
unclear.  As explained above, many algorithms for identifying
influential source spreaders use network topology, and so we expect the
latter also to be a promising source for identifying influential
brokers.

Several algorithms for identifying structural hole spanners~\cite{Lou13:Mining,lin2021structural,xu2017efficient}
that bridge different communities have also been proposed.  Structural
hole spanners are defined as nodes whose removal from a network causes
communities to become disconnected~\cite{xu2017efficient,Lou13:Mining}.
Empirical studies of tweet diffusion on
social media have shown that inter-community diffusion of a tweet increases its final
cascade size~\cite{Bakshy12:Role,Weng13:Virality,tsugawa2019empirical}, which suggests that structural hole spanners
may be related to influential brokers.
However, the relationship between structural hole spanners and influential
brokers has not been investigated.
A simple way to find structural hole spanners is to extract nodes with
high betweenness as used in~\cite{goyal2007structural}.
We compare influential brokers and central nodes based on betweenness,
which are expected to be structural hole spanners, and we examine the
relationship between the concepts of structural hole spanners and brokers.

Few studies have quantified a user's influence as a broker,
but a notable exception is that by 
Bhowmick et al.~\shortcite{Bhowmick19:temporal}, who proposed an algorithm called
SmartInf for identifying influencers from past retweet cascades and 
evaluated the identified
influencers' power as brokers; that study suggests that past retweet
cascades are useful for identifying influential
brokers. In contrast, the effectiveness of
network topology for identifying brokers has not been explored.  We
follow Bhowmick et al.~\shortcite{Bhowmick19:temporal} and examine the usefulness of network embeddings and
traditional centrality measures obtained from network topology for
identifying brokers.

\section{Preliminaries}
\label{sec:pre}
\subsection{Notation}

A social network is represented as a directed graph $G=(V,E)$, where $V$
is a set of nodes representing social media users and $E$ is a set of
links representing the relationships among those users. Link
$(u,v) \in E$ represents the fact that user $u$ follows user $v$.

A sequence of retweets (reposts) and the original tweet (post) is referred to as
an information diffusion cascade. Furthermore, 
$U_c=\{u_1^c, u_2^c, \dots, u_{n^c}^c\}$ is the set of users who 
retweet the
original tweet of cascade $c$, and $T_c=\{t_{u_1}^c, t_{u_2}^c, \dots,t_{u_{n^c}}^c \}$
is the set of timestamps of the retweets in cascade $c$.
Here, $u_i^c$ is the user who posts the $i$-th retweet in cascade $c$, 
$t_{u_i}^c$ is the timestamp of the retweet posted by user $u_i$ in cascade $c$, and $n^c$
is the cascade size of $c$. The user who post the
original tweet of $c$ is denoted as $u_0^c$, and the timestamp of the original tweet
of $c$ is denoted as $t_0^c$.
A set of diffusion cascades among the social media users $V$ is denoted
as $D=\{c\}$, and a set of cascades initiated by node $v$
is denoted as $C_v =\{c \mid c \in D, u_0^c=v \}$.
The set of users who retweet in cascade $c$ after user $v$
retweets in cascade $c$ is denoted as $R_{c}^v=\{u_i^c \mid t_{u_i}^c
> t_{v}^c\}$.

\subsection{Definitions of Influence}

We define influencer scores of each user as both a source spreader and
a broker.

\paragraph{{\bf Source spreader score:}} An influential source spreader is a
user who can spread their own tweets to many other users.  Thus, the source
spreader score of user $u$ is defined as the number of users who retweet
user $u$'s tweets and is given by
\begin{eqnarray}
{\mathcal S}_u=\left|\bigcup_{c \in C_u} U_c \right|.
\end{eqnarray}
This metric measures the popularity of user $u$'s tweets and was used in
previous studies to evaluate the influence of nodes in social
networks~\cite{pei2014,zhang2016identifying} and for influence maximization
problems~\cite{panagopoulos2020influence,Kempe03:Maximizing}.

\paragraph{{\bf Broker score:}} An influential broker is a user who can spread other users' tweets to many other
users.  Thus, following~\cite{Bhowmick19:temporal} and analogously to the source spreader score, the broker score of user $u$ is defined as the number of users who post retweets after 
user $u$'s retweets and is given by
\begin{eqnarray}
{\mathcal B}_u= \left|\bigcup_{c \in D} R_c^u \right|. \label{eq:broker}
\end{eqnarray}
This metric is intended to measure the impact of user $u$'s retweets on the popularity
of the retweeted tweets~\cite{Bhowmick19:temporal}. 
Since it is
difficult to directly measure the impact of each retweet on the future
popularity of the retweeted tweet, we assume that the users who
participate in many large-scale cascades at their early stages are influential
brokers.

\section{Methodology}
\label{sec:method}
\subsection{Datasets}
We use three social media datasets that we refer to as the Twitter Japan, Twitter Nepal~\cite{Bhowmick19:temporal}, and Digg~\cite{hogg2012social} datasets, which contain both
a social network of social media users and information diffusion cascades among them.
The basic statistics of these datasets are given
in Table~\ref{tab:stat}.  
\begin{itemize}
 \item {\bf Twitter Japan:} The Twitter Japan dataset contains a
who-follows-whom network of Japanese Twitter users and their tweets and
retweets posted during January 2014.  This dataset was collected in our
previous study~\cite{tsugawa2019empirical,tsugawa2018identifying,Tsugawa17:Retweets}.
\item {\bf Twitter Nepal:} The Twitter Nepal
dataset~\cite{Bhowmick19:temporal} contains tweets and retweets about the 2015 Nepal
earthquake, as well as a who-follows-whom network among the users involved in
those tweets and retweets.  
\item {\bf Digg:} The Digg dataset~\cite{hogg2012social} contains posts,
their {\em votes}, and a who-follows-whom network among the users.
In the Digg social media, users can make posts (called stories in Digg) and
also vote on other users' posts. A voted post is shown in the
timelines of followers of the voted user. Thus, the sequence of a story
and votes to it is regarded as a cascade~\cite{lerman2010information}.  As given in
Table~\ref{tab:stat}, only 0.15\% of users made any
posts, with most users only voting, and so source spreader scores can be
      calculated for only a few users. Therefore, we used the Digg dataset
not for identifying influential source spreaders but only for
      identifying influential brokers.
\end{itemize}

\begin{table}[tb]
 \centering
\caption{Basic statistics of the datasets.}
\label{tab:stat}
 \resizebox{.99\columnwidth}{!}{\begin{tabular}{l|r|r|r}
\toprule
Datasets & Twitter Japan & Twitter Nepal& Digg\\ \hline
 Num. users& 351,759&273,222 & 279,631\\
 Num. cascades& 8,419,432&26,424 & 3,553\\
 Num. retweets& 29,587,972& 521,938& 3,018,197\\
 Num. source users &256,549 & 18,983& 431\\
\bottomrule
\end{tabular}}
\end{table}

Our three datasets come from different domains, which is expected to be
useful for examining how the characteristics of influencers differ
across social media services, languages, and topics.  Our datasets cover
two social media services (i.e., Twitter and Digg) and three different
languages (i.e., English, Nepali, and Japanese).  Moreover, the Twitter
Nepal dataset contains topic-specific cascades, whereas the Digg and
Twitter Japan datasets contain non-topic-specific cascades.  By using
these three different types of datasets, we examine the characteristics
of influencers in different domains.

\subsection{Centrality}

We extract influential brokers as well as influential source spreaders from the three
datasets, and we examine the characteristics of the brokers and source spreaders using
centrality measures.  We use traditional popular centrality measures
including degree centrality~\cite{Freeman79:Centrality}, closeness
centrality~\cite{Freeman79:Centrality}, betweenness
centrality~\cite{Freeman79:Centrality}, $k$-core
index~\cite{seidman1983,Dorogovtsev06:k}, and
PageRank~\cite{Brin98:PageRank}.  

From social network $G$ representing
who-follows-whom relationships among social media users, we extract the top
$p$\% of central nodes based on the centrality measures.  We also
extract the top $p$\% of users based on the source spreader and broker scores
as influential source spreaders and brokers, respectively.  We then
examine the overlap among the extracted influencers  (i.e., source
spreaders and brokers) and central nodes.
More specifically, we calculate $p$\% overlap scores~\cite{pei2014,tsugawa2018identifying,Borgatti06:robustness} between
influencers
and central nodes as
\begin{eqnarray}
 {\rm Overlap}_p= \frac{\left|T_p^{\rm inf} \cap T_p^{\rm cent}\right|}{\left|T_p^{\rm inf}\right|},
\end{eqnarray}
where $T_p^{\rm inf}$ and $T_p^{\rm cent}$ are the sets of top-$p$\% influencers and central nodes, respectively.
Previous studies~\cite{chen2012,lu2016vital} have shown that influential source spreaders
tend to have high centrality; thus, overlap between influential source
spreaders and central nodes is expected to be high.
In contrast, whether influential brokers have high
centrality is not yet clear.

\subsection{Node Embedding}

We also examine how influential brokers can be characterized by node
embeddings~\cite{Rossi18:WWW,Rossi18:TKDE,goyal2018graph,cui2018survey}.
A node embedding is a latent low-dimensional vector representation of a
node in a network, and a node embedding technique learns such vector
representations of nodes from only the topological structure of a given
network~\cite{goyal2018graph,cui2018survey}.  While traditional
centrality measures are human-crafted features that are defined explicitly
using human knowledge about central nodes, node embeddings are features
learned from the network data alone without explicitly using human
knowledge about the node characteristics.  Node embeddings have been
shown to be effective for several
tasks~\cite{goyal2018graph,cui2018survey}, which motivates us to examine
the effectiveness of learned
embeddings of nodes for characterizing influencers.

Among several options for node embedding techniques~\cite{goyal2018graph,cui2018survey}, we use 
DeepGL~\cite{Rossi18:WWW,Rossi18:TKDE}.  While with most node embeddings it is
difficult to interpret the meaning of each dimension of the obtained
vectors, DeepGL produces interpretable vector representations of
nodes~\cite{Rossi18:WWW,Rossi18:TKDE,fujiwara2020interpretable}.  Using
interpretable embeddings, we try to understand the influencers'
characteristics that are not captured by traditional centrality
measures.

In what follows, we briefly introduce DeepGL; see~\cite{Rossi18:TKDE} for the details.  
DeepGL uses the base features $\vb{x}$ of
each node.  The base features of a node are represented as a vector, and each
dimension can be any feature, such as a traditional centrality measure of
the node or a node attribute.  DeepGL learns the vector
representation of each node from its base features $\vb{x}$ and those
of its neighbors by using a relational function $f$.
A relation function is a combination of relational feature operators that can be applied to
a base feature.  A relational feature operator obtains a value from base
feature values of one-hop neighbors of a target node.  Examples of the
relational feature operator include mean, sum, and max.  In a directed
network, in-neighbor, out-neighbor, and total-neighbor can be defined,
and the operators for in, out, and total neighbors are denoted as
$\Phi^-_S$, $\Phi^+_S$, and $\Phi_S$, respectively.  $S$ is a summary function that returns a
single value from multiple values, such as sum, mean, and max.  For
instance, $\Phi^-_{\rm mean}(x)$ is a relational operator that
calculates the mean $x$ of the in-neighbors of a node.  In DeepGL,
each dimension of the obtained representation vector of a node is defined as
the combination of base features $x$ and relational operators.
For instance, each dimension could be $\Phi_{\rm mean}(\rm{degree})$,
which is the mean of the degree of neighboring nodes. 
More-complex features can be obtained, such as $\Phi_{\rm mean} \circ
\Phi_{\rm max}(\rm{degree})$, which for node $v$ is calculated as follows.
First, for each neighbor $u$ of node $v$, the maximum degree among
$u$'s neighbors is obtained. Then, the mean of the maximum values for node $v$'s
neighbors is obtained.
DeepGL can learn such complex features of nodes from a network in an
unsupervised way by combining base features and relational functions.

\subsection{Predicting Influencers}

We examine the effectiveness of the node embeddings obtained with DeepGL
for identifying influential brokers. 
We conducted experiments on identifying influential brokers from the
node embeddings using a supervised machine-learning framework.
For comparison purposes, we also conducted experiments on identifying
influential source spreaders.
The task here is to identify the top-$p$\% influential brokers (or source
spreaders) from the DeepGL embeddings.
We obtained the DeepGL embeddings from each social network $G$.
We also obtained broker scores (or source spreader
scores) for all nodes.
Then, we annotated nodes with top-$p$\% scores as {\em influencers}, and others as {\em non-influencers}.
A fraction $q$ of nodes in each network $G$ are used as training
data, and the others are used as test data.
For the training data, class labels (i.e., influencers or
non-influencers) are available for model training.
We used LightGBM~\cite{ke2017lightgbm} as a classifier, and for each network we
trained models of identifying influential brokers.  LightGBM was chosen because it can
produce feature importance of the obtained model and is known to be
effective for machine learning tasks using tabular data~\cite{ke2017lightgbm}.
To cope with highly imbalanced class labels (i.e., there are far more
non-influencers than influencers), we subjected the training data to
downsampling and obtained balanced training data, where the numbers of
non-influencers and influencers are the same.
Moreover, 80\% of the downsampled training data were used for model
training, and 20\% of the training data were used for hyperparameter
tuning using Optuna~\cite{akiba2019optuna}.  The loss function was binary
logarithmic loss.
Note that the test data were still imbalanced.
The parameter configurations for DeepGL are summarized in
Table~\ref{tab:config}.
We used the DeepGL implementation in~\cite{fujiwara2020interpretable}.\footnote{\url{https://github.com/takanori-fujiwara/cnrl}}
By applying DeepGL to each social network, we obtained embedding
vectors of nodes in an unsupervised way.
For each setting, we performed 10 experiments and obtained average
scores for classification accuracy.

\begin{table}[tb]
 \centering
\caption{Parameter configurations for DeepGL.}
\label{tab:config}
\begin{tabular}{l|l}
\toprule
Base features & degree, betweenness, closeness\\
& PageRank, $k$-core \\
Relational functions & sum, max, mean \\
$\lambda$ & 0.9\\
Ego network distance & 5 \\
Transform method & log binning \\
\bottomrule
\end{tabular}
\end{table}

\section{Results}
\label{sec:result}

\subsection{Overlap Among Influential Brokers, Source Spreaders, and Central Nodes}

\begin{figure}[tbp]
\centering
\subfloat[Twitter Japan]{\includegraphics[width=.84\columnwidth]{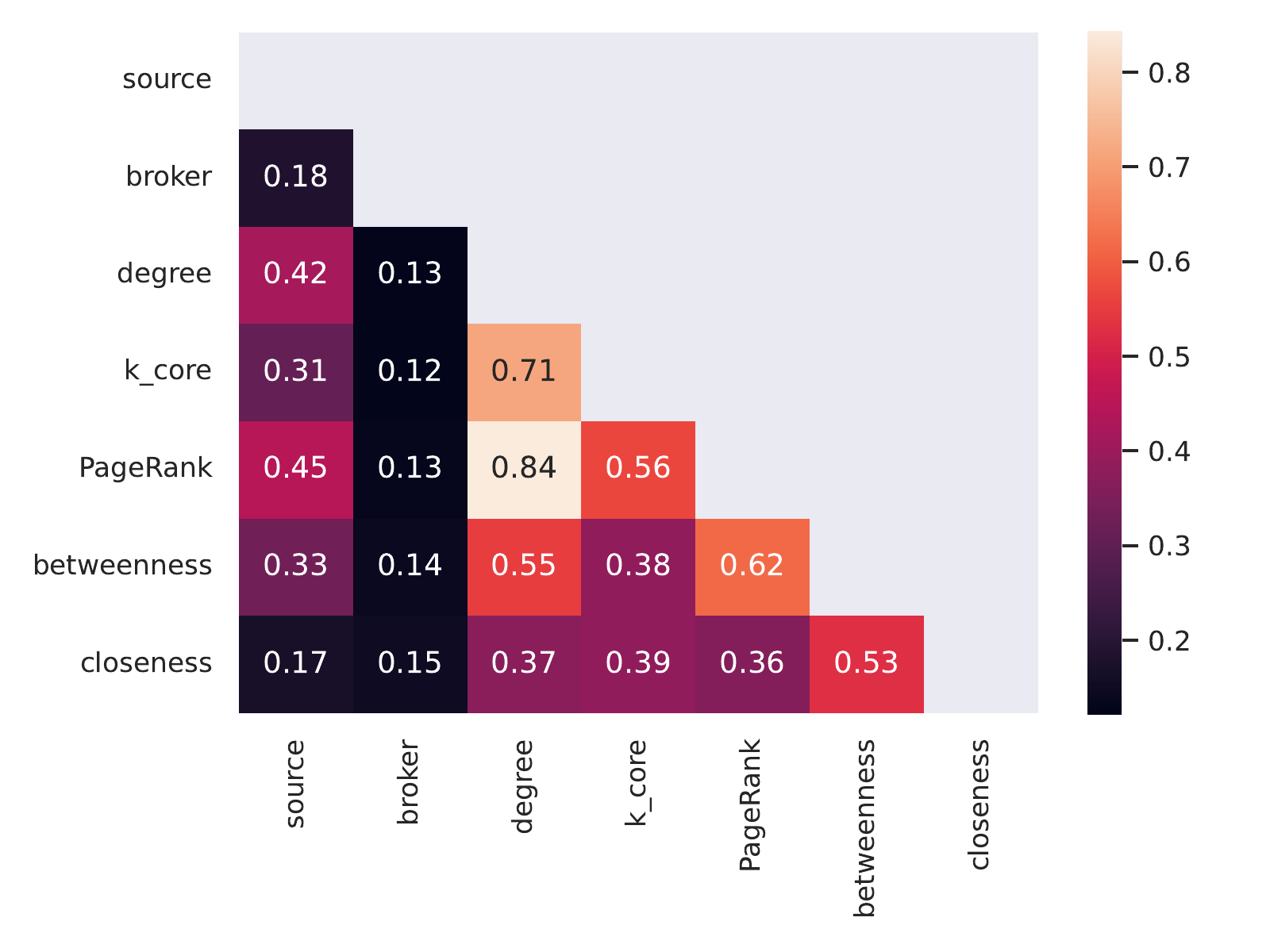}}\\
\subfloat[Twitter Nepal]{\includegraphics[width=.84\columnwidth]{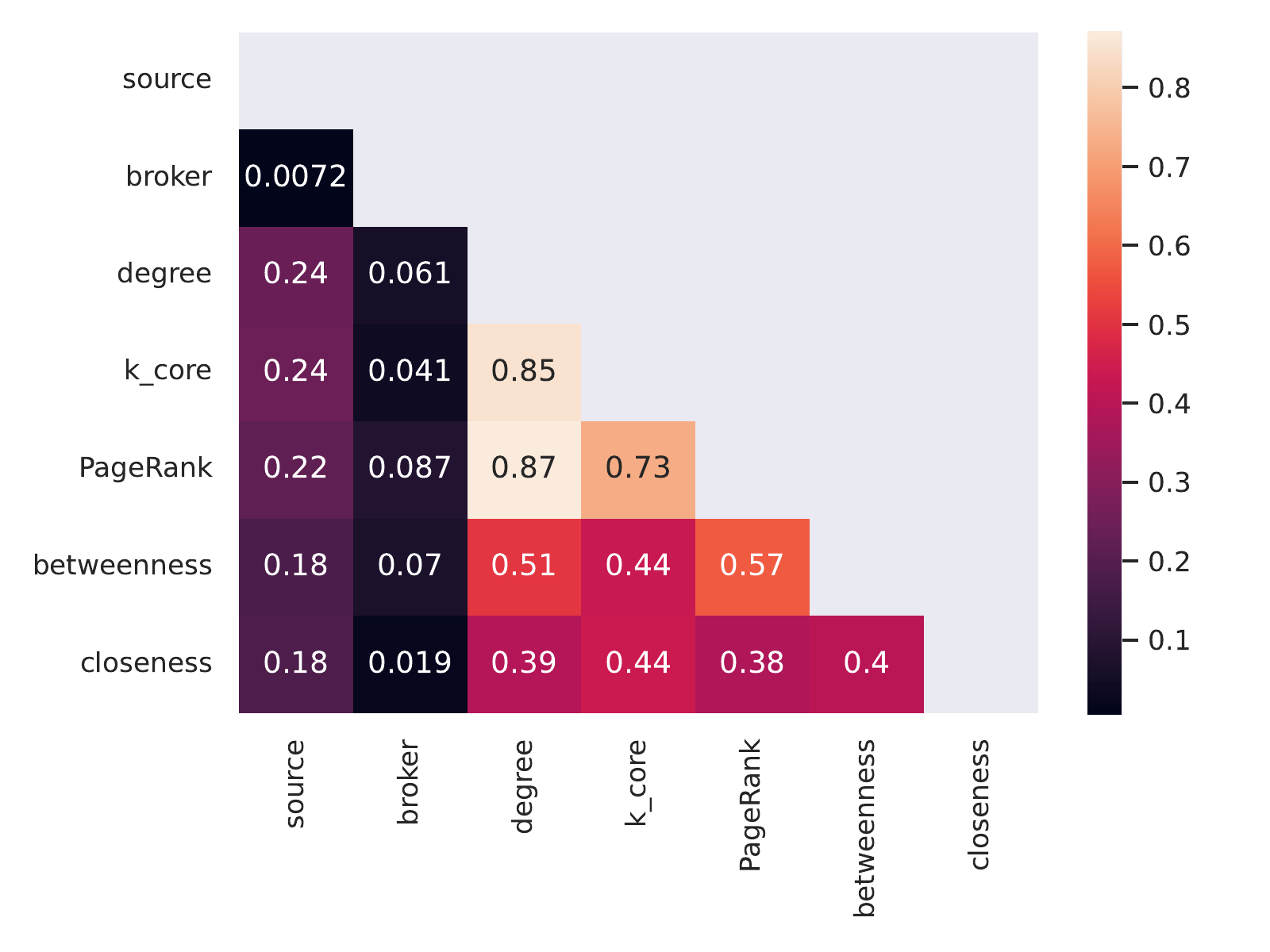}} \\
\subfloat[Digg]{\includegraphics[width=.84\columnwidth]{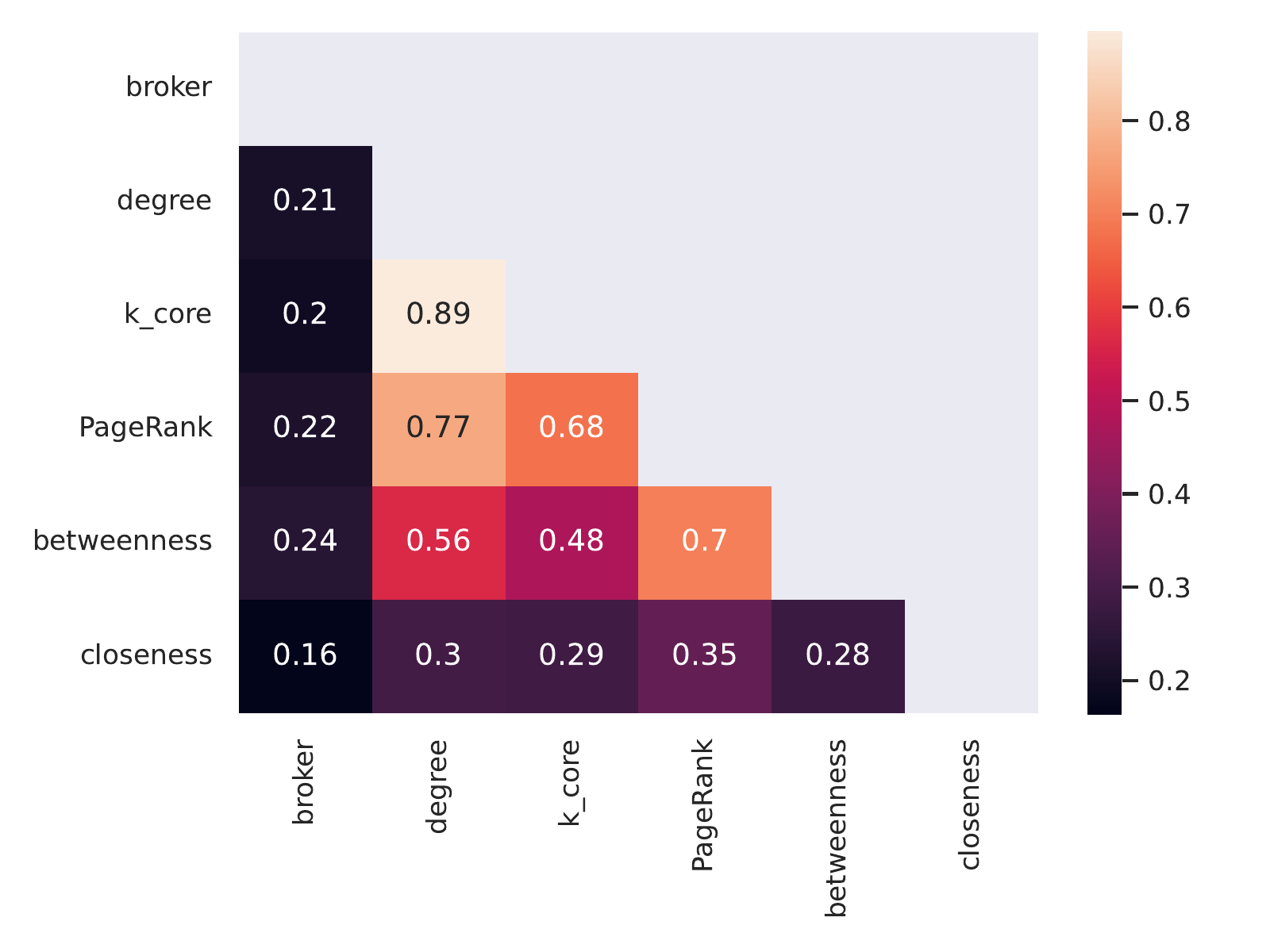}}
    \caption{Confusion matrices for influencers and central
 nodes. Overlap scores between source spreaders and brokers are
 low.  The overlaps between central nodes and brokers 
are low for all datasets, but the overlaps with source spreaders are
 relatively high.}
    \label{fig:heatmap}
\end{figure}

We first examine the characteristics of
influential brokers by comparing them with influential source spreaders
and central nodes ({\bf RQ1} and {\bf RQ2}).  Figure~\ref{fig:heatmap} shows confusion
matrices of 10\% overlap scores among influential brokers, source
spreaders, and central nodes.  Note that influential source spreaders were not extracted from the Digg data because the number of source users was too small (see Table~\ref{tab:stat}).

Figure~\ref{fig:heatmap} shows that overlaps between source spreaders and
brokers are low.  For the Twitter Japan dataset, the
overlap score is 0.18, and for the Twitter Nepal dataset the score is less than
0.01.
These results indicate that influential source spreaders who can
widely disseminate their own messages and brokers who
can widely disseminate other users' messages are different.
Namely, it is suggested that influential brokers who catch the trending topics
early do not have high influence to spread their own messages.

Looking at the overlaps between brokers and central
nodes, we find that the overlap scores are generally low for all
datasets.  The Digg dataset has the highest overlap among the datasets,
but the overlap scores are only about 0.2. For the Twitter Japan and Twitter Nepal
datasets, the overlap scores are 0.1--0.15 and less than 0.1,
respectively.  This indicates that influential brokers cannot be 
characterized well by traditional centrality measures, and a simple heuristic that extracts nodes with high centrality measures~\cite{chen2012,lu2016vital} is not a good means of finding influential brokers.
In contrast, overlaps between source spreaders and central nodes are
relatively high for both the Twitter Japan and Twitter Nepal datasets.
This is consistent with previous studies~\cite{chen2012,lu2016vital}.

\paragraph{{\bf Summary of answers to RQ1 and RQ2:}} The overlap between
brokers and source spreaders is generally small (0.18 for the Twitter Japan
dataset and less than 0.01 for the Twitter Nepal dataset), which suggests that
brokers have different characteristics from source spreaders.  The overlaps between central nodes and brokers are
also small, whereas overlaps with source spreaders
are relatively large, which suggests that using a single centrality
measure is effective for identifying source spreaders but not
brokers.

\subsection{Difference between Influential Brokers and Source Spreaders}

\begin{figure}[tb]
\centering

\subfloat[Twitter
 Japan]{\includegraphics[width=.8\columnwidth]{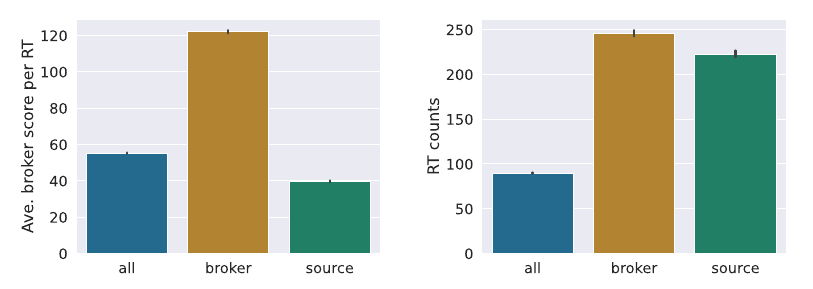}}\\
\subfloat[Twitter Nepal]{\includegraphics[width=.8\columnwidth]{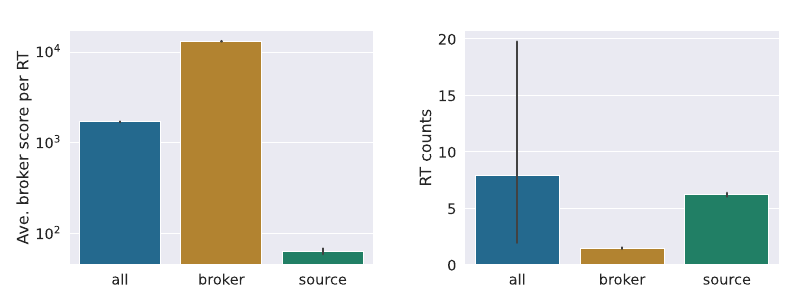}}\\
\subfloat[Digg]{\includegraphics[width=.8\columnwidth]{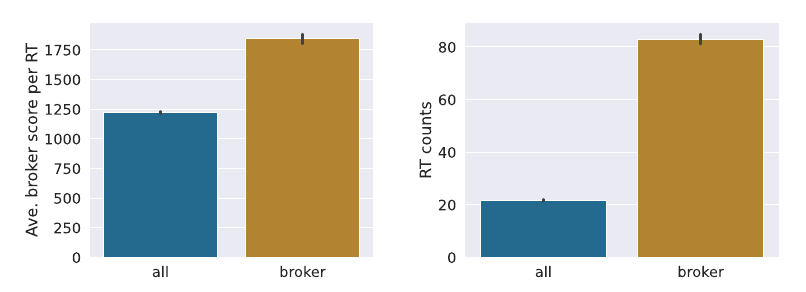}}
    \caption{Comparison of average broker score per retweet and the
 number of retweets posted by users among
 influential source spreaders, influential brokers, and all users.  A tweet retweeted by an influential broker tends to spread more
 widely than does a tweet retweeted by an influential source spreader. }
    \label{fig:bar_ave}
\end{figure}

Furthermore, we investigate the difference between influential brokers and
influential source spreaders.  To characterize a social media user, we investigate the impact of their retweet on the subsequent retweets of other users.
To quantify the impact of a retweet posted by user $u$ on the subsequent
retweets, we define the average broker score
per retweet of user $u$ as ${\mathcal B}_u/r_u$, where $r_u$
is the number of retweets posted by user $u$.  Note that ${\mathcal
B}_u$ is the broker score of user $u$ defined as Eq.~(\ref{eq:broker}).
We compare the average broker score per retweet among influential
brokers, influential source spreaders, and all users.  We also compare
the average number of retweets posted by influential
brokers, influential source spreaders, and all users.
Figure~\ref{fig:bar_ave} shows the results.

These results clearly show the difference between influential brokers and
influential source spreaders.  Figure~\ref{fig:bar_ave} shows that the
average broker score per retweet of influential brokers is
considerably higher than that of influential source spreaders.  These
results show that a tweet retweeted by an influential broker tends to
spread more widely than does a tweet retweeted by an influential source
spreader. Figure~\ref{fig:bar_ave} also shows that the numbers of retweets posted by influential source spreaders tend to be either more than or comparable to the numbers of retweets posted by ordinary users in the Twitter datasets. In contrast, the average broker
score per retweet of influential source spreaders is lower than that of 
ordinary users.  This confirms that although influential source
spreaders' original tweets tend to spread widely, tweets retweeted by
them do not.  Note that the causal relationship between user's
involvement in a retweet cascade and its future cascade size is still unclear from the data.  Namely, it is unclear
whether (i) retweets posted by influential brokers actually influence other
users' retweets or (ii) influential brokers just participate in the early
stages of diffusion of viral tweets.  However, even though the
causality is unclear, it is true that influential brokers have unique
characteristics that differ from those of other users.

\subsection{Predicting Brokers and Source Spreaders}

Next, we examine the accuracy of predicting influential brokers from the
DeepGL embedding features and centrality features ({\bf RQ3}).
Table~\ref{tab:acc-inter} gives the precision, recall, and F$_1$ scores of
identifying the top-10\% and top-5\% influential brokers.  The training data
were 20\% of the nodes, and the other 80\% were used as test data.
The results for the model using only traditional centrality measures as
features and the model using the DeepGL embedding features are given.

\begin{table*}
\centering
\caption{Prediction accuracy for influential brokers. Models using
 DeepGL features achieve considerably higher accuracy than do models
 using only centrality measures.}
\label{tab:acc-inter}
 \begin{tabular}{r|rrr||rrr||rrr}
  \toprule
  \multicolumn{10}{c}{Predicting top-10\% brokers} \\ \hline
& \multicolumn{3}{c||}{Twitter Japan} & \multicolumn{3}{c||}{Twitter Nepal} &
  \multicolumn{3}{c}{Digg} \\ 
model & precision & recall & F1 & precision & recall & F1 & precision &
  recall & F1 \\ \hline
centrality & 0.20 & 0.72 & 0.31 & 0.33 & 0.89 & 0.48 & 0.20 & 0.75 & 0.32 \\
embedding & {\bf 0.26} & {\bf 0.80} & {\bf 0.40} & {\bf 0.53} & {\bf
  0.96} & {\bf 0.68} & {\bf 0.22} & {\bf 0.81} & {\bf 0.35} \\
  \bottomrule
  \multicolumn{10}{c}{ }\\
\toprule
  \multicolumn{10}{c}{Predicting top-5\% brokers} \\ \hline
& \multicolumn{3}{c||}{Twitter Japan} & \multicolumn{3}{c||}{Twitter Nepal} &
  \multicolumn{3}{c}{Digg} \\ 
model & precision & recall & F1 & precision & recall & F1 & precision &
  recall & F1 \\ \hline
centrality & 0.08 & 0.69 & 0.14& 0.15 & 0.85 & 0.25& 0.11 & 0.72 & 0.19\\
embedding & \bf{0.15} & {\bf 0.81} & {\bf 0.25}& {\bf 0.26} & {\bf 0.97}
  & {\bf 0.42}& \bf {0.13} & {\bf 0.80} & {\bf 0.23}\\
  \bottomrule
 \end{tabular}
\end{table*}

\begin{table}
\centering
\caption{Prediction accuracy for influential source spreaders. The
models using 
 DeepGL features and those using only centrality measures have similar accuracy levels.}
\label{tab:acc-source}
  \resizebox{.98\columnwidth}{!}{\begin{tabular}{r|rrr||rrr}
  \toprule
  \multicolumn{7}{c}{Predicting top-10\% source spreaders} \\ \hline
& \multicolumn{3}{c||}{Twitter Japan} & \multicolumn{3}{c}{Twitter Nepal} \\ 
model & precision & recall & F1 & precision & recall & F1  \\ \hline
centrality & 0.28 & 0.80 & 0.42 & 0.11 & 0.60 & 0.19   \\
embedding & {\bf 0.30} & {\bf 0.82}  & {\bf 0.44} & {\bf 0.12} & {\bf
  0.69} &  {\bf 0.21}  \\
  \bottomrule
  \multicolumn{7}{c}{ }\\
\toprule
  \multicolumn{7}{c}{Predicting top-5\% source spreaders} \\ \hline
& \multicolumn{3}{c||}{Twitter Japan} & \multicolumn{3}{c}{Twitter Nepal} \\ 
model & precision & recall & F1 & precision & recall & F1  \\ \hline
centrality & 0.17 & 0.80 & 0.28& {\bf 0.09} & 0.59 & 0.15  \\
embedding & {\bf 0.19} & {\bf 0.83} & {\bf 0.31}& {\bf 0.09} & {\bf
  0.71} & {\bf 0.16} \\
  \bottomrule
 \end{tabular}}
\end{table}

These results show that the prediction models using the DeepGL embedding features achieve higher accuracy than the models using only centrality features.  Particularly for the Twitter Japan and Twitter Nepal datasets, the F$_1$ scores of the
embedding model are respectively 0.11 and 0.17 higher than those of the centrality
model for the task of predicting the top-5\% brokers, which indicates the
effectiveness of the DeepGL embeddings for identifying influential brokers.
For the Digg dataset, the embedding features are less effective than for the Twitter
datasets, but the embedding model achieves higher accuracy than the
centrality model.

For comparison, next we examine the effectiveness of the DeepGL
embeddings for identifying influential source spreaders
(Table~\ref{tab:acc-source}).  The training data were 20\% of the nodes,
and the other 80\% were used as test data.  In contrast to the previous
results, these results show that using the DeepGL embeddings has little
effect on identifying influential source spreaders.  There is little
difference in the prediction accuracy of the models using only
traditional centrality measures and the models using the DeepGL
embeddings.  This indicates that the complex features obtained with
DeepGL are not effective for identifying source spreaders.  
Moreover, combining Tables~\ref{tab:acc-inter} and~\ref{tab:acc-source}, we find that the accuracy of predicting influential brokers is comparable or even higher than that of predicting influential source spreaders.

Note that the obtained accuracy scores are not sufficiently high
for practical use and should be improved in future
research.  For instance, a recent study~\cite{ye2021celebrities}
identified influential users in Sina Weibo with an 
F$_1$ score of approximately 0.8 using state-of-the-art machine learning
techniques such as node embeddings and graph neural networks. 
Although our definition of influential users differs from that used by Ye, Liu, and Pan~\shortcite{ye2021celebrities}, their approaches can still be applied to the task of broker identification. Our results are useful as the first step toward identifying influential brokers, but it will be necessary to construct prediction models with better accuracy and practicality.

\paragraph{{\bf Summary of answer to RQ3:}} The accuracy of predicting the top-10\% influential brokers using the DeepGL embedding features is 0.35--0.68, which is comparable with the accuracy of predicting influential source spreaders.  Using node embeddings as features is suggested to be a more effective approach than using traditional centrality measures.

\subsection{Feature Importance}

Next, we investigate the important features that contribute to the influencer
prediction in the embedding models, and we examine the characteristics
of brokers and source spreaders.  
We obtained feature importance scores [known as the Gini importance~\cite{louppe2013understanding}]
from the constructed models.
Tables~\ref{tab:imp-inter} and~\ref{tab:imp-source} give the top-five
important features based on the importance score in the embedding models
for the tasks of predicting the top-10\% influential brokers and source
spreaders, respectively.  
Here, we show the importance of a model selected randomly from the 10 constructed ones, but
we confirmed that the important features were consistent across the 10 models.

\begin{table}[tbp]
\centering
 \caption{Top-five importance scores of features in the embedding models
 for broker prediction. Complex features using multiple relational
 feature operators have higher importance.  As base features, betweenness,
 $k$-core, and PageRank are used in the important features.}
\label{tab:imp-inter}
 \resizebox{.98\columnwidth}{!}{\begin{tabular}{lrr}
  \toprule
 \multicolumn{3}{c}{Twitter Japan} \\ \hline
  relational function & base feature&importance   \\ \hline
$\Phi^{-}_{\rm mean} \circ \Phi^{-}_{\rm mean} \circ \Phi^{-}_{\rm mean} \circ \Phi^{-}_{\rm mean}$ & betweenness&2430.3 \\ 
$\Phi^{-}_{\rm mean} \circ \Phi^{-}_{\rm mean} \circ \Phi^{-}_{\rm mean}$ & $k$-core &1306.3 \\ 
$\Phi^{}_{\rm mean} \circ \Phi^{}_{\rm mean} \circ \Phi^{-}_{\rm max} \circ \Phi^{-}_{\rm max}$ & $k$-core&1061.7 \\ 
$\Phi^{-}_{\rm mean} \circ \Phi^{-}_{\rm mean} \circ \Phi^{-}_{\rm mean}
				 \circ \Phi^{-}_{\rm max}$ & $k$-core &960.5 \\ 
$\Phi^{-}_{\rm mean} \circ \Phi^{-}_{\rm mean} \circ \Phi^{-}_{\rm mean} \circ \Phi^{-}_{\rm mean}$ & PageRank&895.9 \\ 
  \bottomrule
\toprule
 \multicolumn{3}{c}{Twitter Nepal} \\ \hline
$\Phi^{}_{\rm mean} \circ \Phi^{-}_{\rm mean} \circ \Phi^{-}_{\rm max}$ & PageRank&5469.8 \\ 
$\Phi^{-}_{\rm max} \circ \Phi^{-}_{\rm mean} \circ \Phi^{}_{\rm mean} \circ \Phi^{-}_{\rm max}$ & PageRank&4609.1 \\ 
$\Phi^{-}_{\rm max} \circ \Phi^{-}_{\rm mean} \circ \Phi^{-}_{\rm max}
				 $ & PageRank &4199.7 \\ 
$\Phi^{}_{\rm mean} \circ \Phi^{-}_{\rm max} \circ \Phi^{}_{\rm mean} \circ \Phi^{-}_{\rm mean}$ & PageRank&3883.1 \\ 
$\Phi^{}_{\rm mean} \circ \Phi^{}_{\rm mean} \circ \Phi^{}_{\rm mean}\circ \Phi^{-}_{\rm max}$ & PageRank&2842.5 \\ 
  \bottomrule
\toprule
 \multicolumn{3}{c}{Digg} \\ \hline
$\Phi^{-}_{\rm max} \circ \Phi^{}_{\rm mean} \circ \Phi^{}_{\rm mean} $ & betweenness&2813.2 \\ 
- &betweenness &1621.3 \\ 
- & PageRank&1029.7 \\ 
$\Phi^{-}_{\rm max} \circ \Phi^{}_{\rm mean} \circ \Phi^{-}_{\rm mean}$ & betweenness&1025.7 \\ 
$\Phi^{-}_{\rm max} \circ \Phi^{}_{\rm mean} \circ \Phi^{+}_{\rm max}$ &
				     degree &905.0 \\ 
  \bottomrule
 \end{tabular}}
\end{table}

\begin{table}[tbp]
\centering
 \caption{Top-five importance scores of features in the embedding models
 for source spreader prediction. In contrast to the broker prediction
 models, centrality measures have high importance
 in the source spreader prediction models.}
\label{tab:imp-source}
 \resizebox{.98\columnwidth}{!}{\begin{tabular}{lrr}
  \toprule
 \multicolumn{3}{c}{Twitter Japan} \\ \hline
  relational function & base feature&importance 
  \\ \hline
- & PageRank&12260.7 \\ 
- & degree&2213.9 \\ 
$\Phi^{}_{\rm mean} \circ \Phi^{-}_{\rm mean} \circ \Phi^{}_{\rm mean}$ & betweenness&1381.8 \\ 
$\Phi^{-}_{\rm max} \circ \Phi^{-}_{\rm mean} \circ \Phi^{-}_{\rm max} \circ \Phi^{-}_{\rm mean}$ & $k$-core&1058.5 \\ 
$\Phi^{-}_{\rm sum} \circ \Phi^{-}_{\rm mean} \circ \Phi^{-}_{\rm mean}$ &closeness &853.1 \\ 
  \bottomrule
\toprule
 \multicolumn{3}{c}{Twitter Nepal} \\ \hline
$\Phi^{-}_{\rm mean} \circ \Phi^{+}_{\rm max} \circ \Phi^{}_{\rm mean} \circ \Phi^{}_{\rm mean}$ & PageRank&268.5 \\ 
$\Phi^{}_{\rm mean} \circ \Phi^{-}_{\rm max} \circ \Phi^{}_{\rm mean} \circ \Phi^{}_{\rm mean}$  &betweenness &220.7 \\ 
- &betweenness &213.2 \\ 
$\Phi^{-}_{\rm max} \circ \Phi^{}_{\rm mean} \circ \Phi^{}_{\rm mean} \circ \Phi^{-}_{\rm max}$ & degree&202.1 \\ 
- & $k$-core &174.6 \\ 
 \bottomrule
 \end{tabular}}
\end{table}

Table~\ref{tab:imp-inter} shows that for the Twitter datasets, traditional centrality
measures are not included in the top-five features.
Complex embedding features that use three or four feature operators are shown to
be effective features for identifying influential brokers.
Features obtained from $l$-feature operators incorporate
features of nodes in $l$-hop neighbors.  Thus, features with a large
number of feature operators incorporate complex higher-order structural
features of nodes.  These results suggest that such complex features are useful for identifying
influential brokers in Twitter networks.
For the Digg dataset, although the DeepGL embeddings are included in the
top-five features, traditional centrality measures are also included.
Moreover, the numbers of feature operators used in the DeepGL features
are smaller than those for the Twitter datasets.
For the Digg dataset, compared with the Twitter datasets, the effectiveness
of the DeepGL features is suggested to be limited, which is consistent
with the results of the prediction experiments (Table~\ref{tab:acc-inter}).

In contrast to the broker prediction, Table~\ref{tab:imp-source} shows that traditional centrality
measures are included as top-five important features for the source spreader
prediction tasks.
Particularly for the Twitter Japan dataset, PageRank has the highest
importance score.  
The DeepGL embedding features are also included in the top-five important
features, but the importance scores are comparable or lower than
traditional centrality measures.
These results confirm that for the source spreader
prediction tasks, the DeepGL embeddings are not so effective, and
traditional centrality measures are sufficient.

These results also show that many features based on betweenness
centrality, PageRank, and $k$-core are included in the top-ranked lists.
However, effective features are different for the tasks and datasets.
From these results, we cannot find universal characteristics of
influential brokers and source spreaders across different topics
and user sets. 

\subsection{Transferability}

The results in the previous subsection raise a new question about the
transferability of the constructed models.  We obtained quite different
results for each different dataset, which suggests that a prediction
model learned from one dataset may not be effective on other datasets.
Therefore, we evaluate the prediction accuracy of a constructed model on one
dataset when applying it to other datasets.  We train a model using a
{\em source-domain} dataset, then we evaluate its accuracy 
on a {\em target-domain} dataset; the training procedure is the
same as that in the previous subsection.  Since each dimension of the feature vector
must be consistent between the training (i.e., source-domain) data and
the test (i.e., target-domain) data, we transfer the
DeepGL node embeddings from the source domain to the target domain when
using the DeepGL embedding model. DeepGL can produce consistent
node embedding vectors across different networks by using its inductive
learning framework~\cite{Rossi18:WWW,Rossi18:TKDE}.

Table~\ref{tab:transfer-broker} compares the F$_1$ scores of the task of 
identifying the top-10\% influential brokers for different combinations of
source- and target-domain datasets, and Table~\ref{tab:transfer-source} does
the same for the task of identifying the top-10\% influential source spreaders.
These results show that when the source and target
domains are different, the F$_1$ scores are considerably lower than when
the source and target domains are the same, suggesting that
transferring a trained influencer identification model to different
domains is difficult. 
Our results also suggest that it is more difficult to transfer a broker 
identification model than a source-spreader one.
The models trained on different domains achieve only poor
accuracy, particularly in the broker identification task.
As suggested by the results in the previous subsection, traditional
centrality measures are more useful for influential source-spreader 
identification than for influential broker identification, which 
might be the cause of the difference in transferability.
Overall, the results suggest that models for identifying
influencers should be trained for each domain to achieve high prediction
accuracy.  

\begin{table}
\centering
\caption{Comparison of F$_1$ scores of the models among different
 combinations of source and target domain datasets when predicting
 top-10\% brokers.}
\label{tab:transfer-broker}
 \resizebox{.99\columnwidth}{!}{ \begin{tabular}{r|rrr||rrr||rrr}
  \toprule
Target& \multicolumn{3}{c||}{Twitter Japan} &
				  \multicolumn{3}{c||}{Twitter Nepal} & 
  \multicolumn{3}{c}{Digg} \\ 
Source & Japan & Nepal & Digg& Japan & Nepal & Digg & Japan
   & Nepal & Digg\\ \hline
centrality &0.31  &0.03  & 0.18 & 0.06& 0.48 & 0.18 & 0.10 & 0.02 & 0.32 \\
embedding & 0.40 & 0.05 &0.17  & 0.11 & 0.68 & 0.14 & 0.17 & 0.05 &0.35  \\
  \bottomrule
 \end{tabular}}
\end{table}

\begin{table}
\centering
\caption{Comparison of F$_1$ scores of the models among different
 combinations of source and target domain datasets when predicting
 top-10\% source spreaders.}
\label{tab:transfer-source}
 \begin{tabular}{r|rr||rr}
  \toprule
Target& \multicolumn{2}{c||}{Twitter Japan} &
				  \multicolumn{2}{c}{Twitter Nepal}  \\ 
Source & Japan & Nepal & Japan & Nepal \\ \hline
centrality & 0.42 &0.22  & 0.18 & 0.19 \\
embedding & 0.44 & 0.31 &0.19  &  0.21 \\
  \bottomrule
 \end{tabular}
\end{table}

\subsection{Effects of Amount of Training Data}

\begin{figure}[tb]
  \centering 
\subfloat[Brokers]{\includegraphics[width=.49\columnwidth]{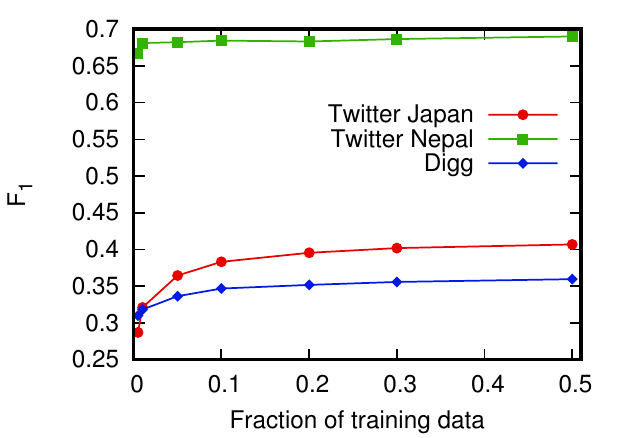}}
\subfloat[Source spreaders]{\includegraphics[width=.49\columnwidth]{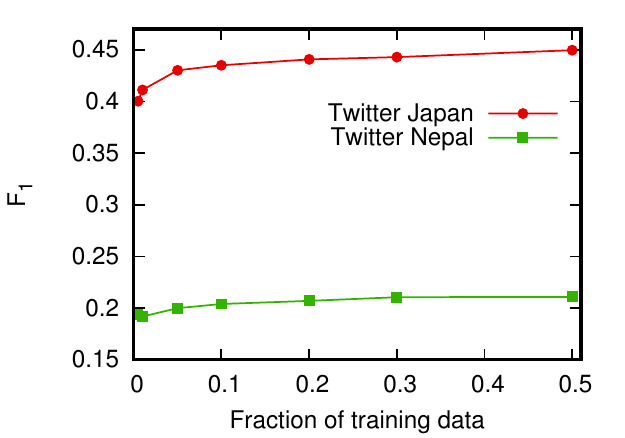}} 
\caption{Effects of amount of training data when predicting top-10\%
 influencers. Using 5\% training data
 achieves F$_1$ scores that are similar to those with 50\% training data.}
  \label{fig:train}
\end{figure}

Finally, we investigate how the amount of training data affects the
prediction accuracy of the constructed models.  Figure~\ref{fig:train}
shows F$_1$ scores of the embedding models for each dataset while
changing the fraction of training data. The task here was predicting the top-10\%
influencers, and the source and target domains were the same.
Figure~\ref{fig:train} shows
that for all the datasets, the F$_1$ scores with 5\% training data are almost the
same as those with 50\% training data.  This suggests that
5\% training data are enough to learn the model of identifying
influencers.  The results suggest that a small amount of training data is enough for obtaining influencer prediction models. 
While results in the previous subsection suggest that influencer identification models
should be trained for each domain, the results in this subsection suggest that the training
cost for each dataset is not large, which is preferable in practice.

\section{Discussion}
\label{sec:discuss}

\subsection{Implications}
Our results show that the characteristics of influential brokers are different from those of influential source spreaders, which suggests that algorithms for identifying source spreaders cannot be used directly to identify brokers.  Although designing centrality measures has been effective for identifying influential source spreaders~\cite{lu2016vital,Li14:Identifying,zhang2016identifying}, our results suggest that such an approach is not effective for identifying brokers.  

Our results show that the unique characteristics of influential brokers
differ from those of influential source spreaders.  Tweets retweeted by
influential brokers tend to spread to many users, whereas tweets
retweeted by influential source spreaders do not.  Note that the causal
relationship between a broker's involvement in a retweet cascade and its
future cascade size is unclear.  
One possible explanation is
that a retweet posted by an influential broker facilitates other users'
retweets, which affects the future popularity of the tweet.  The other
explanation is that influential brokers retweet tweets that will
be popular in the future in
their early stages of retweet diffusion.  In both cases, we reason that
identifying influential brokers is useful.  In the former case,
influential brokers would be useful in viral marketing: a company could
ask influential brokers to spread the tweets posted by its
accounts so that those tweets spread widely.  In the latter case,
influential brokers would be useful for knowing which content 
will be popular in the future; also, identifying such users could help 
limit the spread of unwanted information by asking them not to do so.

Our results also show that while a single centrality measure has poor
predictive power in identifying influential brokers, considerable
accuracy is achieved by combining multiple centrality measures and using
node embeddings.  This indicates that human-crafted centrality features
fail to capture the characteristics of influential brokers, but social
network structure contains important information for finding influential
brokers.  We therefore expect that to further improve the accuracy of
predicting influential brokers, node representation learning as used in
this paper will be promising.  Although we used DeepGL embeddings to
obtain an interpretable representation, as used
in~\cite{ye2021celebrities}, other representation learning techniques including graph neural networks~\cite{kipf2017semi,Wu20:Comprehensive} could be used to improve the prediction accuracy.

\subsection{Limitations}

This study has some limitations, which we discuss below along with future research directions.  First, the effectiveness of other social network
features for identifying brokers should be investigated.  There are many
options for node embedding techniques~\cite{goyal2018graph,cui2018survey}, and ones
that are suitable for identifying brokers should be explored in
future research.  While using betweenness centrality alone is shown to
be ineffective, we are interested in investigating the effectiveness
of other techniques for identifying structural hole spanners~\cite{Lou13:Mining,lin2021structural,xu2017efficient}
for identifying influential brokers.

Second, the prediction accuracy should be improved for practical use.
Although finding top influencers is a difficult task, we expect
that there is room for improvement in the prediction accuracy of the
models.  As suggested in~\cite{Bhowmick19:temporal}, past log data of
diffusion cascades are expected to be a useful source.  Therefore, a model
that incorporates both topological structure and diffusion cascades is
expected to achieve higher accuracy.
Moreover, using features of tweet contents such as linguistic features
obtained from language models and like counts is also
expected to be useful, and 
analyzing specific cases in which influential brokers play an important
role (e.g., information diffusion regarding state-sponsored troll accounts~\cite{zannettou2019disinformation}) may also provide useful insights.

Third, the applicability of the trained model to other users should be
further investigated.  
Our results show that the trained models in this paper cannot be
applied to different domains; 
this is not surprising because the three datasets differ considerably
in terms of languages, cultures, topics, and
social media platforms, but we should investigate further the
transferability of our approach.
For instance, a model
trained on the Twitter Japan dataset might be applicable to different
Japanese user sets. 
Also, we are interested
in methods for transferring a pre-trained model to other
domains~\cite{qiu2020gcc} for influencer identification, and 
clarifying a method for constructing a transferable influencer identification
model is important future work.

Fourth, as discussed already, whether retweets posted by influential
brokers do actually influence the retweeting behaviors of other users 
remains unclear.  Clarifying a causal relation between one user's retweet
and other users' retweets is challenging but important future work, 
and a possible way to tackle this problem would be to conduct a field 
experiment to ask social media users to retweet some
specific tweets, and then compare their future diffusion patterns.

\section{Conclusion}
\label{sec:conclusion}
In this paper, we tackled the problem of identifying influential brokers who can spread other users' messages to many users on social media.  Using three social media datasets, we investigated the characteristics of influential brokers by comparing them with influential source spreaders and central nodes.  Our results showed that most of the influential source spreaders are not influential brokers (and vice versa).  We also showed that overlap between central nodes and influential brokers was less than 15\% on the two Twitter datasets, which suggests that a heuristic that extracts highly central nodes as brokers is not a good approach.  We conducted experiments of identifying influential brokers by using node embedding features obtained with DeepGL.  Our results showed that models using DeepGL embeddings achieved F$_1$ scores of 0.35--0.68, which is a similar level of accuracy to that of identifying influential source spreaders.  Moreover, models using DeepGL embeddings achieved higher accuracy than did those using only centrality measures, which indicates the effectiveness of DeepGL embeddings for identifying brokers.  Our results showed the effectiveness of using network topology for identifying influential brokers, as well as the limitations of using traditional centrality measures.

\section{Acknowledgments}
 This work was supported by JSPS KAKENHI Grant No. 20H04172 and 22K11990.

\bibliography{bib/influence_short}

\begin{thebibliography}{65}
\providecommand{\natexlab}[1]{#1}
\providecommand{\url}[1]{\texttt{#1}}
\providecommand{\urlprefix}{URL }
\expandafter\ifx\csname urlstyle\endcsname\relax
  \providecommand{\doi}[1]{doi:\discretionary{}{}{}#1}\else
  \providecommand{\doi}{doi:\discretionary{}{}{}\begingroup
  \urlstyle{rm}\Url}\fi

\bibitem[{Akiba et~al.(2019)Akiba, Sano, Yanase, Ohta, and
  Koyama}]{akiba2019optuna}
Akiba, T.; Sano, S.; Yanase, T.; Ohta, T.; and Koyama, M. 2019.
\newblock Optuna: A next-generation hyperparameter optimization framework.
\newblock In \emph{Proc. KDD'19}, 2623--2631.

\bibitem[{Al-Garadi et~al.(2018)Al-Garadi, Varathan, Ravana, Ahmed, Mujtaba,
  Khan, and Khan}]{al2018analysis}
Al-Garadi, M.~A.; Varathan, K.~D.; Ravana, S.~D.; Ahmed, E.; Mujtaba, G.; Khan,
  M. U.~S.; and Khan, S.~U. 2018.
\newblock Analysis of online social network connections for identification of
  influential users: Survey and open research issues.
\newblock \emph{ACM Computing Surveys (CSUR)} 51(1): 1--37.

\bibitem[{Araujo, Neijens, and Vliegenthart(2017)}]{araujo2017getting}
Araujo, T.; Neijens, P.; and Vliegenthart, R. 2017.
\newblock Getting the word out on Twitter: The role of influentials,
  information brokers and strong ties in building word-of-mouth for brands.
\newblock \emph{International Journal of Advertising} 36(3): 496--513.

\bibitem[{Bakshy et~al.(2011)Bakshy, Hofman, Mason, and
  Watts}]{Bakshy11:Everyone's}
Bakshy, E.; Hofman, J.~M.; Mason, W.~A.; and Watts, D.~J. 2011.
\newblock Everyone's an Influencer: {Quantifying} Influence on {Twitter}.
\newblock In \emph{Proc. WSDM'11}, 65--74.

\bibitem[{Bakshy et~al.(2012)Bakshy, Rosenn, Marlow, and
  Adamic}]{Bakshy12:Role}
Bakshy, E.; Rosenn, I.; Marlow, C.; and Adamic, L. 2012.
\newblock The role of social networks in information diffusion.
\newblock In \emph{Proc. WWW'12}, 519--528.

\bibitem[{Banerjee, Jenamani, and Pratihar(2020)}]{banerjee2020survey}
Banerjee, S.; Jenamani, M.; and Pratihar, D.~K. 2020.
\newblock A survey on influence maximization in a social network.
\newblock \emph{Knowledge and Information Systems} 62(9): 3417--3455.

\bibitem[{Bhowmick et~al.(2019)Bhowmick, Gueuning, Delvenne, Lambiotte, and
  Mitra}]{Bhowmick19:temporal}
Bhowmick, A.~K.; Gueuning, M.; Delvenne, J.-C.; Lambiotte, R.; and Mitra, B.
  2019.
\newblock Temporal sequence of retweets help to detect influential nodes in
  social networks.
\newblock \emph{IEEE Transactions on Computational Social Systems} 6(3):
  441--455.

\bibitem[{Borgatti, Carley, and Krackhardt(2006)}]{Borgatti06:robustness}
Borgatti, S.~P.; Carley, K.~M.; and Krackhardt, D. 2006.
\newblock On the robustness of centrality measures under conditions of
  imperfect data.
\newblock \emph{Social Networks} 28(2): 124--136.

\bibitem[{Bouguessa and Romdhane(2015)}]{Bouguessa15:Identifying}
Bouguessa, M.; and Romdhane, L.~B. 2015.
\newblock Identifying authorities in online communities.
\newblock \emph{ACM Transactions on Intelligent Systems and Technology (TIST)}
  6(3): 1--23.

\bibitem[{Brin and Page(1998)}]{Brin98:PageRank}
Brin, S.; and Page, L. 1998.
\newblock The anatomy of a large-scale hypertextual {Web} search engine.
\newblock \emph{Computer Networks and ISDN Systems} 30(1): 107--117.

\bibitem[{Bucur(2020)}]{bucur2020top}
Bucur, D. 2020.
\newblock Top influencers can be identified universally by combining classical
  centralities.
\newblock \emph{Scientific Reports} 10(1): 1--14.

\bibitem[{Budak, Agrawal, and El~Abbadi(2011)}]{Budak11:Limiting}
Budak, C.; Agrawal, D.; and El~Abbadi, A. 2011.
\newblock Limiting the spread of misinformation in social networks.
\newblock In \emph{Proc. WWW'11}, 665--674.

\bibitem[{Burt(2000)}]{burt2000network}
Burt, R.~S. 2000.
\newblock The network structure of social capital.
\newblock \emph{Research in Organizational Behavior} 22: 345--423.

\bibitem[{Cha et~al.(2010)Cha, Haddadi, Benevenuto, and
  Gummadi}]{Cha10:Measuring}
Cha, M.; Haddadi, H.; Benevenuto, F.; and Gummadi, K. 2010.
\newblock Measuring user influence in twitter: The million follower fallacy.
\newblock In \emph{Proc. ICWSM'10}, 10--17.

\bibitem[{Chen et~al.(2012)Chen, L{\"u}, Shang, Zhang, and Zhou}]{chen2012}
Chen, D.; L{\"u}, L.; Shang, M.-S.; Zhang, Y.-C.; and Zhou, T. 2012.
\newblock Identifying influential nodes in complex networks.
\newblock \emph{Physica A: Statistical Mechanics and Its Applications} 391(4):
  1777--1787.

\bibitem[{Cui et~al.(2018)Cui, Wang, Pei, and Zhu}]{cui2018survey}
Cui, P.; Wang, X.; Pei, J.; and Zhu, W. 2018.
\newblock A survey on network embedding.
\newblock \emph{IEEE Transactions on Knowledge and Data Engineering} 31(5):
  833--852.

\bibitem[{Domingos and Richardson(2001)}]{Domingos01:Mining}
Domingos, P.; and Richardson, M. 2001.
\newblock Mining the network value of customers.
\newblock In \emph{Proc. KDD'01}, 57--66.

\bibitem[{Dorogovtsev, Goltsev, and Mendes(2006)}]{Dorogovtsev06:k}
Dorogovtsev, S.~N.; Goltsev, A.~V.; and Mendes, J. F.~F. 2006.
\newblock K-core organization of complex networks.
\newblock \emph{Physical Review Letters} 96(4): 040601.

\bibitem[{Freeman(1979)}]{Freeman79:Centrality}
Freeman, L.~C. 1979.
\newblock Centrality in social networks conceptual clarification.
\newblock \emph{Social Networks} 1(3): 215--239.

\bibitem[{Fujiwara et~al.(2020)Fujiwara, Zhao, Chen, Yu, and
  Ma}]{fujiwara2020interpretable}
Fujiwara, T.; Zhao, J.; Chen, F.; Yu, Y.; and Ma, K.-L. 2020.
\newblock Interpretable contrastive learning for networks.
\newblock \emph{arXiv preprint arXiv:2005.12419} .

\bibitem[{Goyal and Ferrara(2018)}]{goyal2018graph}
Goyal, P.; and Ferrara, E. 2018.
\newblock Graph embedding techniques, applications, and performance: A survey.
\newblock \emph{Knowledge-Based Systems} 151: 78--94.

\bibitem[{Goyal and Vega-Redondo(2007)}]{goyal2007structural}
Goyal, S.; and Vega-Redondo, F. 2007.
\newblock Structural holes in social networks.
\newblock \emph{Journal of Economic Theory} 137(1): 460--492.

\bibitem[{Grover and Leskovec(2016)}]{grover2016node2vec}
Grover, A.; and Leskovec, J. 2016.
\newblock node2vec: Scalable feature learning for networks.
\newblock In \emph{Proc. KDD'16}, 855--864.

\bibitem[{Hogg and Lerman(2012)}]{hogg2012social}
Hogg, T.; and Lerman, K. 2012.
\newblock Social dynamics of {Digg}.
\newblock \emph{EPJ Data Science} 1(1): 1--26.

\bibitem[{Hu et~al.(2012)Hu, Liu, Wei, Wu, Stasko, and Ma}]{hu2012breaking}
Hu, M.; Liu, S.; Wei, F.; Wu, Y.; Stasko, J.; and Ma, K.-L. 2012.
\newblock Breaking news on {Twitter}.
\newblock In \emph{Proc. CHI'12}, 2751--2754.

\bibitem[{Katz and Lazarsfeld(1955)}]{Katz55:personal}
Katz, E.; and Lazarsfeld, P.~F. 1955.
\newblock \emph{Personal influence: the part played by people in the flow of
  mass communications.}
\newblock Free Press.

\bibitem[{Ke et~al.(2017)Ke, Meng, Finley, Wang, Chen, Ma, Ye, and
  Liu}]{ke2017lightgbm}
Ke, G.; Meng, Q.; Finley, T.; Wang, T.; Chen, W.; Ma, W.; Ye, Q.; and Liu,
  T.-Y. 2017.
\newblock {LightGBM}: A highly efficient gradient boosting decision tree.
\newblock In \emph{Proc. NIPS'17}, 3146--3154.

\bibitem[{Kempe, Kleinberg, and Tardos(2003)}]{Kempe03:Maximizing}
Kempe, D.; Kleinberg, J.~M.; and Tardos, E. 2003.
\newblock Maximizing the spread of influence through a social network.
\newblock In \emph{Proc. KDD'03}, 137--146.

\bibitem[{Kipf and Welling(2017)}]{kipf2017semi}
Kipf, T.~N.; and Welling, M. 2017.
\newblock Semi-Supervised Classification with Graph Convolutional Networks.
\newblock In \emph{Proc. ICLR'17}, 1--14.

\bibitem[{Lerman and Ghosh(2010)}]{lerman2010information}
Lerman, K.; and Ghosh, R. 2010.
\newblock Information contagion: An empirical study of the spread of news on
  {Digg} and {Twitter} social networks.
\newblock In \emph{Proc. ICWSM'10}, 90--97.

\bibitem[{Li et~al.(2014{\natexlab{a}})Li, Peng, Li, Sun, Li, and
  Xu}]{li2014social}
Li, J.; Peng, W.; Li, T.; Sun, T.; Li, Q.; and Xu, J. 2014{\natexlab{a}}.
\newblock Social network user influence sense-making and dynamics prediction.
\newblock \emph{Expert Systems with Applications} 41(11): 5115--5124.

\bibitem[{Li et~al.(2014{\natexlab{b}})Li, Zhou, L{\"u}, and
  Chen}]{Li14:Identifying}
Li, Q.; Zhou, T.; L{\"u}, L.; and Chen, D. 2014{\natexlab{b}}.
\newblock Identifying influential spreaders by weighted {LeaderRank}.
\newblock \emph{Physica A: Statistical Mechanics and its Applications} 404:
  47--55.

\bibitem[{Li et~al.(2018)Li, Fan, Wang, and Tan}]{Li18:Influence}
Li, Y.; Fan, J.; Wang, Y.; and Tan, K.-L. 2018.
\newblock Influence maximization on social graphs: A survey.
\newblock \emph{IEEE Transactions on Knowledge and Data Engineering} 30(10):
  1852--1872.

\bibitem[{Lin et~al.(2021)Lin, Zhang, Gong, Chen, Oksanen, and
  Ding}]{lin2021structural}
Lin, Z.; Zhang, Y.; Gong, Q.; Chen, Y.; Oksanen, A.; and Ding, A.~Y. 2021.
\newblock Structural hole theory in social network analysis: A review.
\newblock \emph{IEEE Transactions on Computational Social Systems} 1--16.

\bibitem[{Liu-Thompkins and Rogerson(2012)}]{liu2012rising}
Liu-Thompkins, Y.; and Rogerson, M. 2012.
\newblock Rising to stardom: An empirical investigation of the diffusion of
  user-generated content.
\newblock \emph{Journal of Interactive Marketing} 26(2): 71--82.

\bibitem[{Lou and Tang(2013)}]{Lou13:Mining}
Lou, T.; and Tang, J. 2013.
\newblock Mining structural hole spanners through information diffusion in
  social networks.
\newblock In \emph{Proc. WWW'13}, 825--836.

\bibitem[{Louppe et~al.(2013)Louppe, Wehenkel, Sutera, and
  Geurts}]{louppe2013understanding}
Louppe, G.; Wehenkel, L.; Sutera, A.; and Geurts, P. 2013.
\newblock Understanding variable importances in forests of randomized trees.
\newblock In \emph{Proc. NIPS'13}, 431--439.

\bibitem[{L{\"u} et~al.(2016)L{\"u}, Chen, Ren, Zhang, Zhang, and
  Zhou}]{lu2016vital}
L{\"u}, L.; Chen, D.; Ren, X.-L.; Zhang, Q.-M.; Zhang, Y.-C.; and Zhou, T.
  2016.
\newblock Vital nodes identification in complex networks.
\newblock \emph{Physics Reports} 650: 1--63.
\newblock \doi{10.1016/j.physrep.2016.06.007}.

\bibitem[{L{\"u} et~al.(2011)L{\"u}, Zhang, Yeung, and Zhou}]{lu2011leaders}
L{\"u}, L.; Zhang, Y.-C.; Yeung, C.~H.; and Zhou, T. 2011.
\newblock Leaders in social networks, the delicious case.
\newblock \emph{PloS One} 6(6): e21202.

\bibitem[{Meng et~al.(2018)Meng, Peng, Tan, Liu, Cheng, and
  Bae}]{meng2018diffusion}
Meng, J.; Peng, W.; Tan, P.-N.; Liu, W.; Cheng, Y.; and Bae, A. 2018.
\newblock Diffusion size and structural virality: The effects of message and
  network features on spreading health information on twitter.
\newblock \emph{Computers in Human Behavior} 89: 111--120.

\bibitem[{Morone and Makse(2015)}]{Morone15:CI}
Morone, F.; and Makse, H.~A. 2015.
\newblock Influence maximization in complex networks through optimal
  percolation.
\newblock \emph{Nature} 524(7563): 65--68.

\bibitem[{Panagopoulos, Malliaros, and
  Vazirgianis(2020)}]{panagopoulos2020influence}
Panagopoulos, G.; Malliaros, F.~D.; and Vazirgianis, M. 2020.
\newblock Influence maximization using influence and susceptibility embeddings.
\newblock In \emph{Proc. ICWSM'20}, volume~14, 511--521.

\bibitem[{Pei et~al.(2014)Pei, Muchnik, Andrade~Jr, Zheng, and Makse}]{pei2014}
Pei, S.; Muchnik, L.; Andrade~Jr, J.~S.; Zheng, Z.; and Makse, H.~A. 2014.
\newblock Searching for superspreaders of information in real-world social
  media.
\newblock \emph{Scientific Reports} 4: 5547.

\bibitem[{Qiu et~al.(2020)Qiu, Chen, Dong, Zhang, Yang, Ding, Wang, and
  Tang}]{qiu2020gcc}
Qiu, J.; Chen, Q.; Dong, Y.; Zhang, J.; Yang, H.; Ding, M.; Wang, K.; and Tang,
  J. 2020.
\newblock {GCC}: Graph contrastive coding for graph neural network
  pre-training.
\newblock In \emph{Proc. KDD'20}, 1150--1160.

\bibitem[{Qiu et~al.(2019)Qiu, Dong, Ma, Li, Wang, Wang, and
  Tang}]{qiu2019netsmf}
Qiu, J.; Dong, Y.; Ma, H.; Li, J.; Wang, C.; Wang, K.; and Tang, J. 2019.
\newblock {NetSMF}: Large-scale network embedding as sparse matrix
  factorization.
\newblock In \emph{Proc. WWW'19}, 1509--1520.

\bibitem[{Qiu et~al.(2018)Qiu, Dong, Ma, Li, Wang, and Tang}]{qiu2018network}
Qiu, J.; Dong, Y.; Ma, H.; Li, J.; Wang, K.; and Tang, J. 2018.
\newblock Network embedding as matrix factorization: Unifying {DeepWalk},
  {LINE}, {PTE}, and node2vec.
\newblock In \emph{Proc. WSDM'18}, 459--467.

\bibitem[{Richardson and Domingos(2002)}]{Domingos02:Mining}
Richardson, M.; and Domingos, P. 2002.
\newblock Mining knowledge-sharing sites for viral marketing.
\newblock In \emph{Proc. KDD'02}, 61--70.

\bibitem[{Riquelme and Gonz{\'a}lez-Cantergiani(2016)}]{Riquelme16:Measuring}
Riquelme, F.; and Gonz{\'a}lez-Cantergiani, P. 2016.
\newblock Measuring user influence on {Twitter}: {A} survey.
\newblock \emph{Information Processing \& Management} 52(5): 949--975.

\bibitem[{Rossi, Zhou, and Ahmed(2018{\natexlab{a}})}]{Rossi18:TKDE}
Rossi, R.~A.; Zhou, R.; and Ahmed, N.~K. 2018{\natexlab{a}}.
\newblock Deep inductive graph representation learning.
\newblock \emph{IEEE Transactions on Knowledge and Data Engineering} 32(3):
  438--452.

\bibitem[{Rossi, Zhou, and Ahmed(2018{\natexlab{b}})}]{Rossi18:WWW}
Rossi, R.~A.; Zhou, R.; and Ahmed, N.~K. 2018{\natexlab{b}}.
\newblock Deep inductive network representation learning.
\newblock In \emph{Companion Proc. WWW'18}, 953--960.

\bibitem[{Seidman(1983)}]{seidman1983}
Seidman, S.~B. 1983.
\newblock Network structure and minimum degree.
\newblock \emph{Social Networks} 5(3): 269--287.

\bibitem[{Tang, Shi, and Xiao(2015)}]{Tang15:Influence}
Tang, Y.; Shi, Y.; and Xiao, X. 2015.
\newblock Influence maximization in near-linear time: A martingale approach.
\newblock In \emph{Proc. SIGMOD'15}, 1539--1554.

\bibitem[{Tang, Xiao, and Shi(2014)}]{Tang14:TIM}
Tang, Y.; Xiao, X.; and Shi, Y. 2014.
\newblock Influence maximization: Near-optimal time complexity meets practical
  efficiency.
\newblock In \emph{Proc. SIGMOD'14}, 75--86.

\bibitem[{Tsugawa(2019)}]{tsugawa2019empirical}
Tsugawa, S. 2019.
\newblock Empirical analysis of the relation between community structure and
  cascading retweet diffusion.
\newblock In \emph{Proc. ICWSM'19}, volume~13, 493--504.

\bibitem[{Tsugawa and Kimura(2018)}]{tsugawa2018identifying}
Tsugawa, S.; and Kimura, K. 2018.
\newblock Identifying influencers from sampled social networks.
\newblock \emph{Physica A: Statistical Mechanics and its Applications} 507:
  294--303.

\bibitem[{Tsugawa and Kito(2017)}]{Tsugawa17:Retweets}
Tsugawa, S.; and Kito, K. 2017.
\newblock Retweets as a predictor of relationships among users on social media.
\newblock \emph{PloS One} 12(1): e0170279.

\bibitem[{Weng et~al.(2010)Weng, Lim, Jiang, and He}]{weng2010}
Weng, J.; Lim, E.-P.; Jiang, J.; and He, Q. 2010.
\newblock {TwitterRank}: finding topic-sensitive influential twitterers.
\newblock In \emph{Proc. WSDM'10}, 261--270.

\bibitem[{Weng, Menczer, and Ahn(2013)}]{Weng13:Virality}
Weng, L.; Menczer, F.; and Ahn, Y.-Y. 2013.
\newblock Virality prediction and community structure in social networks.
\newblock \emph{Scientific Reports} 3: 2522.

\bibitem[{Wu et~al.(2020)Wu, Pan, Chen, Long, Zhang, and
  Philip}]{Wu20:Comprehensive}
Wu, Z.; Pan, S.; Chen, F.; Long, G.; Zhang, C.; and Philip, S.~Y. 2020.
\newblock A comprehensive survey on graph neural networks.
\newblock \emph{IEEE Transactions on Neural Networks and Learning Systems}
  32(1): 4--24.

\bibitem[{Xu et~al.(2017)Xu, Rezvani, Liang, Yu, and Liu}]{xu2017efficient}
Xu, W.; Rezvani, M.; Liang, W.; Yu, J.~X.; and Liu, C. 2017.
\newblock Efficient algorithms for the identification of top-$k$ structural
  hole spanners in large social networks.
\newblock \emph{IEEE Transactions on Knowledge and Data Engineering} 29(5):
  1017--1030.

\bibitem[{Yamaguchi et~al.(2010)Yamaguchi, Takahashi, Amagasa, and
  Kitagawa}]{yamaguchi2010turank}
Yamaguchi, Y.; Takahashi, T.; Amagasa, T.; and Kitagawa, H. 2010.
\newblock {TURank}: Twitter user ranking based on user-tweet graph analysis.
\newblock In \emph{Proc. WISE'10}, 240--253.

\bibitem[{Ye, Liu, and Pan(2021)}]{ye2021celebrities}
Ye, W.; Liu, Z.; and Pan, L. 2021.
\newblock Who are the celebrities? Identifying vital users on {Sina} {Weibo}
  microblogging network.
\newblock \emph{Knowledge-Based Systems} 231: 107438.

\bibitem[{Zannettou et~al.(2019)Zannettou, Caulfield, De~Cristofaro,
  Sirivianos, Stringhini, and Blackburn}]{zannettou2019disinformation}
Zannettou, S.; Caulfield, T.; De~Cristofaro, E.; Sirivianos, M.; Stringhini,
  G.; and Blackburn, J. 2019.
\newblock Disinformation warfare: Understanding state-sponsored trolls on
  {Twitter} and their influence on the web.
\newblock In \emph{Companion Proc. WWW'19}, 218--226.

\bibitem[{Zhang et~al.(2016)Zhang, Chen, Dong, and Zhao}]{zhang2016identifying}
Zhang, J.-X.; Chen, D.-B.; Dong, Q.; and Zhao, Z.-D. 2016.
\newblock Identifying a set of influential spreaders in complex networks.
\newblock \emph{Scientific Reports} 6: 27823.

\bibitem[{Zhao et~al.(2020)Zhao, Jia, Huang, Zhou, and Fang}]{zhao2020machine}
Zhao, G.; Jia, P.; Huang, C.; Zhou, A.; and Fang, Y. 2020.
\newblock A machine learning based framework for identifying influential nodes
  in complex networks.
\newblock \emph{IEEE Access} 8: 65462--65471.

\end{thebibliography}


\end{document}